# Stoch-IMC: A Bit-Parallel Stochastic In-Memory Computing Architecture Based on STT-MRAM

**Amir M. Hajisadeghi, Hamid R. Zarandi, Mahmoud Momtazpour\***
Department of Computer Engineering, Amirkabir University of Technology (Tehran Polytechnic)

**Abstract:** In-memory computing (IMC) offloads parts of the computations to memory to fulfill the performance and energy demands of applications such as neuromorphic computing, machine learning, and image processing. Fortunately, the main features that stochastic computing (SC) and IMC share, which are low computation complexity and high bit-parallel computation capability, promise great potential for integrating SC and IMC. In this paper, we exploit this potential by using stochastic computation as an approximation method to present effective in-memory computations with a good trade-off among design parameters. To this end, first, commonly used stochastic arithmetic operations of applications are effectively implemented using the primitive logic gates of the IMC method. Next, the in-memory scheduling and mapping of applications are obtained efficiently by a proposed algorithm. This algorithm reduces the computation latency by enabling intra-subarray parallelism while considering the IMC method constraints. Subsequently, a bit-parallel stochastic IMC architecture, Stoch-IMC, is presented that enables bit parallelization of stochastic computations over memory subarrays/banks. To evaluate Stoch-IMC's effectiveness, various analyses were conducted. Results show average performance improvements of 135.7X and 124.2X across applications compared to binary IMC and related in-memory SC methods, respectively. The results also demonstrate an average energy reduction of 1.5X compared to binary IMC, with limited energy overhead relative to the in-memory SC method. Furthermore, the results reveal average lifetime improvements of 4.9X and 216.3X over binary IMC and in-memory SC methods, respectively, along with high bitflip tolerance.

## 1- Introduction

Modern data-intensive applications like neuromorphic computing, machine learning, and image/video analytics are growing gradually [1,2]. These applications suffer from high data transfer overheads, and conventional computing systems cannot handle them with acceptable throughput and energy efficiency [2,3]. In-memory computing (IMC) is one of the main routes to mitigate the high data transfer demands of the applications [3-5]. The idea is to perform data-intensive computations in memory to minimize the data transfer costs.

Various implementations of the IMC approach have been proposed [6-10]. Among the memristive-based IMC methods, IMPLY [6], MAGIC [7], CRAM [8], and STT-CiM [9] are some of the efficient stateful logics ones. These methods realize the operations in memory by leveraging the memristor's resistive feature and enabling different degrees of parallelism across memory cells which is an essential advantage. However, due to optimal memory design considerations, dark silicon challenges, and manufacturing process cost, most IMC methods are bounded to logical and simple binary arithmetic operations [2,4,11,12]. Subsequently, performing complex binary arithmetic operations such as multiplication, division, and non-linear functions by such IMC methods is inefficient [12-15]. The reason is the necessity to split them into the supported logics and operations of the IMC method, which takes many cycles to complete and is neither latency- nor energy-efficient [12-15].

One promising solution to address this deficiency is to decrease the computational complexity of arithmetic operations by employing stochastic computing (SC), an unconventional approximate computation method that relies on probabilities [16,17]. By reducing the computation complexity, SC provides a trade-off among design parameters such as performance, energy consumption, area, and accuracy [17,18].

This work integrates and innovates the advantages of memristive IMC and SC to present effective in-memory computations: 1) Most memristive IMC methods provide a great parallelism in memory array for computation which can be fully utilized by SC due to the independence of bits in stochastic numbers. Therefore, the operation on one bit can be performed independently of the operation on other bits in a parallel fashion, and the bit-dependency latency of the carry chain in binary computations is removed. 2) Unlike binary computing, many arithmetic operations that are common in neuromorphic and machine learning applications can be highly simplified in the stochastic domain using a low number of logic gates and performed efficiently in memory [17,18]. For instance, multiplication, addition, and subtraction of two stochastic numbers can be performed using one logic gate [16,19]. Also, more complex operations like hyperbolic tangent, exponential functions, and square root can be effectively performed in the stochastic domain using only a few logic gates [16,19,20]. In this regard, simplifying the in-memory operations with SC can greatly reduce computing latency. 3) Random numbers generation (RNG) is an essential component in SC. When implemented with conventional CMOS circuits—typically using linear feedback shift registers or ring oscillators—it incurs high area and energy costs and results in pseudo-random numbers [16, 19]. However, an intrinsic stochastic switching feature in resistive memories can be leveraged to generate stochastic numbers without the need for costly conventional RNG circuits [21]. While this approach reduces area and energy costs, it does not reduce overall latency if the generation step is performed separately from the computation step. This issue can be effectively alleviated by integrating both steps within the memory. 4) Memristive IMCs are constrained by their lifetime as a reliability factor. The endurance of memristors allows for a finite number of accesses. Therefore, due to the greater complexity of binary computations, which necessitates numerous memory accesses, binary memristive IMC becomes more challenging than stochastic one. 5) Bitflip occurrences in stochastic computation, whether due to memristive IMC reliability challenges or other unwanted sources, have a negligible effect on the output value compared to the binary computing domain, especially when they occur in the most significant bits of binary computation. Additionally, in comparison to binary computations, the error propagation probability is considerably lower [15].

---

\* Corresponding author.
E-mail address: momtazpour@aut.ac.ir (Mahmoud Momtazpour)

Although a few works have tried to exploit this match [22-25], they either consider some of the aforementioned advantages or are bound to specific operations. In this work, we fully exploit the great match between memristive IMC and SC, and we present various operations to enhance the performance and energy efficiency of memristive IMC. In this regard, first, efficient implementation of in-memory stochastic arithmetic operations such as adder, multiplier, subtractor, and divider are done. Next, an algorithm is proposed that schedules and maps the stochastic computations of each application in memory taking into account the constraints of the IMC method, and aims to reduce computation latency by enabling intra-subarray parallelism. After that, a stochastic IMC architecture, named Stoch-IMC, is presented, which facilitates bit parallelization of computations over subarrays/banks. Finally, to evaluate the efficiency of this work, the used arithmetic operations in most stochastic applications are assessed, and analyses are conducted at the circuit, architecture, and application levels.

We demonstrate average performance improvements of 135.7X, and 124.2X compared to binary IMC and related in-memory SC research [22], respectively. Additionally, the results indicate an average energy reduction of 1.5X and limited energy overhead compared to binary IMC and research in [22], respectively. Furthermore, evaluations of lifetime and bitflip tolerance demonstrate superiority. The main contributions of this paper are as follows:

- Realizing in-memory binary and stochastic forms of the arithmetic operations that are prevalent in numerous neuromorphic and machine learning applications.
- Proposing an in-memory scheduling and mapping algorithm based on 2T-1MTJ IMC method.
- Presenting Stoch-IMC, a bit-parallel stochastic in-memory computing architecture based on STT-MRAM.
- Conducting analyses at the circuit, architecture, and application levels demonstrates the effectiveness of Stoch-IMC in four stochastic applications.

The rest of the paper is organized as follows: Section 2 provides a brief overview of STT-MRAM, the IMC method, and the stochastic computing preliminaries. Sections 3 and 4 cover related work and the proposed method, including the realization of stochastic arithmetic operations in memory, the scheduling and mapping algorithm, and the proposed stochastic IMC architecture. Section 5 presents the evaluation results, and Section 6 concludes the paper.

## 2- Background

### 2-1- STT-MRAM

STT-MRAM cell consists of an MTJ (magnetic tunnel junction) storage element in series with an NMOS access transistor (1T-1MTJ). It stands out as one of the most promising non-volatile memory (NVM) technologies, thanks to its fast access time, low static energy, great endurance, high density, and resistance to radiation [26]. The MTJ consists of three layers: a MgO barrier layer positioned between two ferromagnetic layers. The outer layers are known as the reference and free layers, depending on their ability to change the direction of the magnetic field. The MTJ is in the parallel state (*P* state with low resistance) when the magnetic field directions of these two layers are the same; if they differ, it is in the anti-parallel state (*AP* state with high resistance). It is possible to assume that *P* and *AP* states correspond to logic values '0' and '1,' respectively. By flowing a current greater than critical current density (i.e., $J_c$, one of the features of the cell), the cell state is altered during a write operation. Additionally, a read operation is carried out by passing a read current through the cell and measuring the MTJ resistance ($I_{read}$ << $I_c$, where $I_c$ is the critical switching current). Figure 1(a) depicts the STT-MRAM cell in the perpendicular structure.

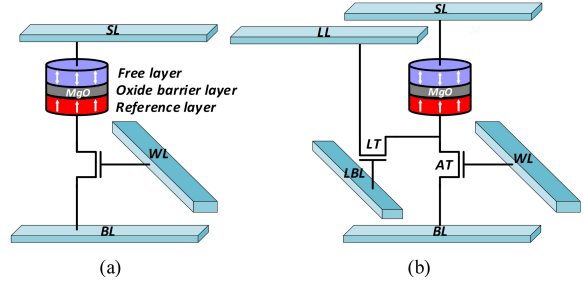

Fig. 1. Structure of a) STT-MRAM, and b) 2T-1MTJ cells.

**2-2- 2T-1MTJ IMC Basis**

The capability for in-memory computation can be integrated into STT-MRAM by incorporating an extra NMOS transistor. The resultant structure is known as CRAM and shown in Fig. 1(b) [8]; however, due to multiple CRAM structures, we refer to it as 2T-1MTJ in-memory computation method. The 2T-1MTJ bit-cell operates in two modes: memory mode and logic mode. The memory mode is activated by appropriately utilizing the word line (*WL*), bit line (*BL*), and source line (*SL*), similar to the operation of STT-MRAM. The logic mode, on the other hand, is activated by using two extra lines, the logic line (*LL*) and the logic bit line (*LBL*), and is accomplished by enabling the logic transistor (*LT*) while disabling the access transistor (*AT*).

For instance, 2T-1MTJ IMC method performs a logic operation, as is represented in Fig. 2(a) and described below. To conduct an intra-row logic operation, consider that the inputs (*in1* and *in2*) are stored in the first two MTJs as resistances, and according to the desired logic function, the output MTJ (*out*) is preset. $V_{SL}$ is then applied to the *SL* lines of the input cells, while the *SL* line of the output cell is grounded. Finally, based on the current flowing through the output MTJ, its resistance may either change or remain the same, generating the output logic. It is important to note that each logic gate has a specific $V_{SL}$ range and output cell preset value, which have been presented in [3, 44].

To clearly illustrate how performing a logic operation using the 2T-1MTJ IMC method, one NAND logic operation with four input combinations is exemplified, and its simulation result is shown in Fig. 2(b). To perform NAND logic, the output cell should be preset to logic '0', and the input cells resistances are set to desired values. It is done in the memory mode of the 2T-1MTJ cell similar to the write operation of STT-MRAM. Thereafter, in the logic mode of the 2T-1MTJ cell, the *WL* and the *LBL*s are set to ground and $V_{dd}$, respectively. Also, $SL_0$, $SL_1$ (related to input cells), and $SL_2$ (related to output cell) will be $V_{SL}$ and ground, respectively. Subsequently, the logic result is stored as a resistance in the output cell.

This logic result can be obtained through a read operation or by observing the output cell state via the $T_{trans}(out)$ signal, as shown in the simulation waveform of Fig. 2(b). The state and transition time of cells can be observed via $T_{trans}$ signal; therefore, $T_{trans}(in1)$, $T_{trans}(in2)$, and $T_{trans}(out)$ can demonstrate the correctness of the NAND logic operation. Finally, it should be noted that although all four input combinations of the NAND logic gate are shown in Fig. 2(b), the preset phase for performing NAND(0,0) was done in the pre-simulation configurations and is not shown in these simulation waveforms.

It is worth mentioning that, unlike IMC-P (peripheral) methods, which perform computations by moving data from the memory array to peripheral circuits (such as sense amplifiers or custom CMOS-based embedded computing units) and then writing it back, 2T-1MTJ IMC method conducts computations directly within the memory array, known as IMC-A (array) methods, thereby eliminating unnecessary data transfers [27]. Additionally, in contrast to analog IMC approaches that compute based on current and voltage characteristics, this method is digital, meaning it processes digital data stored in the memory array, resulting in great computational robustness [28,29].

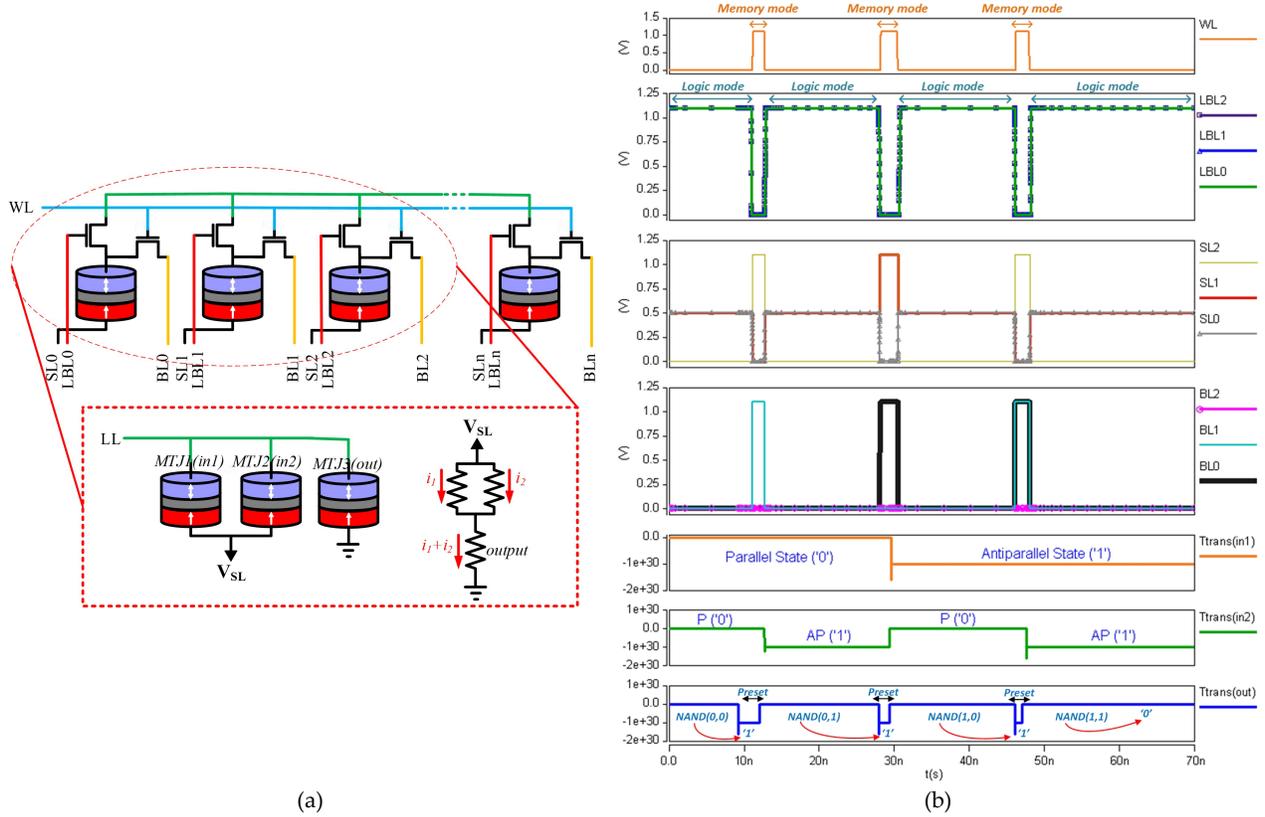

(a) (b)

Fig. 2. An example of logic operation using 2T-1MTJ IMC method, a) cells' structure and connections, and b) simulation waveform of NAND operation (signals *LBL*, *SL*, and *BL* were depicted using different colors and marker styles for better showing).

### 2-3- Stochastic Computing

In SC, a stochastic number (SN) is represented as a sequence of 0s and 1s, known as a bitstream. The value of the SN depends on the encoding mode used. For instance, in unipolar encoding mode, the probability of a bit being '1' in the bitstream corresponds to the value of the SN. Additionally, a binary number with *N*-bit precision can be represented using a $2^N$-bit bitstream [16]. In general, a stochastic processing system comprises three steps [16,30]: 1) stochastic number generation (SNG), 2) stochastic computation, and 3) stochastic to binary conversion. The first step is typically implemented by a comparator with two inputs: desired data and a random number. Every clock cycle, a random number is generated, and the comparator outputs '1' if the input value is greater than the random number; otherwise, it outputs '0.' The intended precision of SN determines the number of bits in the bitstream, which is equal to the overall number of required clock cycles.

Memristors have an intrinsic stochastic switching feature which makes them suitable as a probabilistic element to generate tunable true random numbers and stochastic numbers, eliminating the need for typical costly CMOS-based SNG unit in SC [21,31]. Based on this feature, the switching delay of memristor bit cells exhibits a non-deterministic nature due to inherent physical behaviors, such as random thermal fluctuations of the MTJ in STT-MRAM [21,32]. Therefore, applying a voltage pulse of specific amplitude and duration causes the MTJ state to switch randomly but with a controlled probability. We utilize this intrinsic stochasticity feature in MTJ of STT-MRAM to generate tunable stochastic numbers. In this regard, the cell value is preset to logic '0' (*P*-state), and then a voltage pulse corresponding to the desired stochastic value is applied, leading to the state of the MTJ cell changing with a probability matching the desired value. The corresponding amplitude ($V_P$) and duration ($t_P$) of the applied voltage pulse that impacts the switching probability ($P_{sw}$) of the MTJ can be obtained by [21,32],

$$P_{sw} = 1 - e^{-\frac{t_p}{\tau}} \quad (1)$$

$$\tau = \tau_0 e^{\Delta\left(1 - \frac{V_p}{V_{c0}}\right)} \quad (2)$$

where $\Delta$ represents the MTJ thermal stability factor, and $V_{c0}$ and $\tau_0$ denote the critical switching voltage of the MTJ and the thermal attempt time at 0K, respectively. Figure 3 illustrates the relationship between $P_{sw}$ and $V_p$ for

$t_p$ values ranging from 3$ns$ to 10$ns$ [33]. It shows that the switching probability is proportional to $V_p$ and $t_p$, increasing with the applied voltage pulse amplitude and duration. As an example, by applying a voltage pulse with an amplitude of 310$mV$ and a duration of 4$ns$, switching occurs with a probability of 0.7. Therefore, if this voltage pulse is applied ten times on preset cells, seven cells will change to logic '1' and three cells will remain at logic '0', resulting in a stochastic number equal to 0.7.

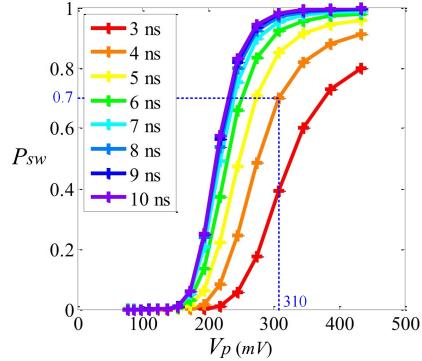

Fig. 3. Relationship between switching probability ($P_{sw}$) and pulse amplitude ($V_P$) of MTJ for varying pulse durations ($t_P$).

In the second step, the considered computation is performed in the SC domain over the provided input bitstreams. In this paper, we represent SNs in unipolar encoding mode which supports data range between 0–1. In unipolar mode, the arithmetic operations are replaced by simple logical operations. The stochastic logic circuit of common arithmetic operations has been shown in Fig. 4. In this regard, the input operands of these operations are independent bitstreams, except for absolute value subtraction which uses correlated input bitstreams. Also, for most applications, $S$ in scaled addition (Fig. 4(a)) is set to 0.5. Finally, in the third step, by counting 1s in the output bitstream, we can get the equivalent binary value of the output SN, which is conventionally achievable by a digital counter. In this paper, we concentrate on the second step, stochastic data computation, which is the most time-consuming one, and the third step is performed similarly to conventional SC schemes.

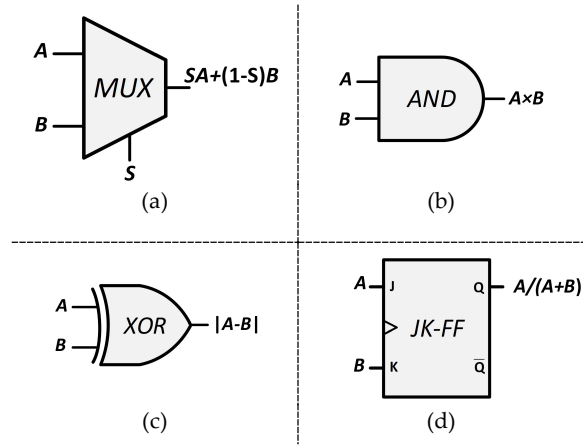

Fig. 4. Stochastic logic circuit of common arithmetic operations a) scaled addition, b) multiplication, c) absolute value subtraction, and d) scaled division in unipolar encoding.

## 3- Related Work

Numerous studies explore the use of SC in various applications, including neural networks and image/video processing. For example, some studies improve the energy and performance efficiency of neural networks by performing a subset of the computations in the stochastic domain [13,14,34-36]. Research [36] presents a Bayesian belief network for heart disaster prediction and a Bayesian inference system for object location application. Additionally, studies [37], and [38] are some of the image processing applications that get significant design parameter improvements by performing stochastic computation.

Nevertheless, there is currently a focus on efficiently integrating SC with memristive IMC in such applications to alleviate their high data transfer overhead. Studies [23], [24], and [25] are among the first attempts that integrate operations such as stochastic full-addition or multiplication by resistive memories. Work [39] extends the use of stochastic in-memory operations and briefly highlights the potential of this integration in data-intensive applications. Moreover, two studies [15], and [33] examine the reliability aspects of integrating SC with memristive IMC. Study [15] explores the reliability challenges of applying SC to memristive IMC, while study [33] evaluates how changes in memory technology impacted the accuracy of stochastic-based in-memory computation.

Research [22] is one of the first well-known digital IMC studies, which comprehensively investigates the potential of embedding stochastic computation and CRAM IMC method and assesses different neuromorphic applications in terms of different design parameters. In [22], the computations for each bit are presented and repeated according to the bitstream length. However, a mechanism for storing the result of each bit in the bitstream to complete all bits' computations has not been provided. This method relies on a single memory array, and a memory architecture to effectively handle SC and IMC integration has not been presented.

## 4- Stoch-IMC: The Proposed Method

Stoch-IMC is proposed to efficiently execute data-intensive approximable applications in memory by utilizing SC. First, common primary stochastic arithmetic operations used in most applications are realized by 2T-1MTJ IMC method. Second, a heuristic algorithm is presented that co-schedules and maps the stochastic computations to memory based on 2T-1MTJ IMC method. Lastly, the Stoch-IMC bit-parallel memory architecture is presented.

### 4-1- In-memory Realization of Stochastic Arithmetic Operations

In order to realize stochastic arithmetic operations in memory, two phases should be done: 1) The first phase is converting a given arithmetic operation to the equivalent logic circuit using supported primitive logic gates provided by IMC method, and 2) The second phase is the in-memory implementation of the equivalent logic circuit. The 2T-1MTJ IMC method supports logic gates such as BUFF, INV, AND, NAND, OR, and NOR. Based on phase one, the equivalent gate-level circuit of common stochastic arithmetic operations plus the square root and exponential operations based on the supported primitive IMC logic gates have been presented in Fig. 5 [16,20]. In this regard, unlike other operations, the two inputs in the stochastic circuit for the absolute value subtraction operation need to be correlated (Fig. 5(c)). Also, Q should be initially set to zero in the scaled division circuit (Fig. 5(d)). The two inputs ($A_1$ and $A_2$) in square root circuit (Fig. 5(e)) have the same value but are independently generated and have different bitstreams, and $C_1$ and $C_2$ are two constant bitstreams. Finally, the exponential, $e^{-cA}$, is implemented based on the fifth order of Maclaurin expansion where $0 < c \leq 1$ [20].

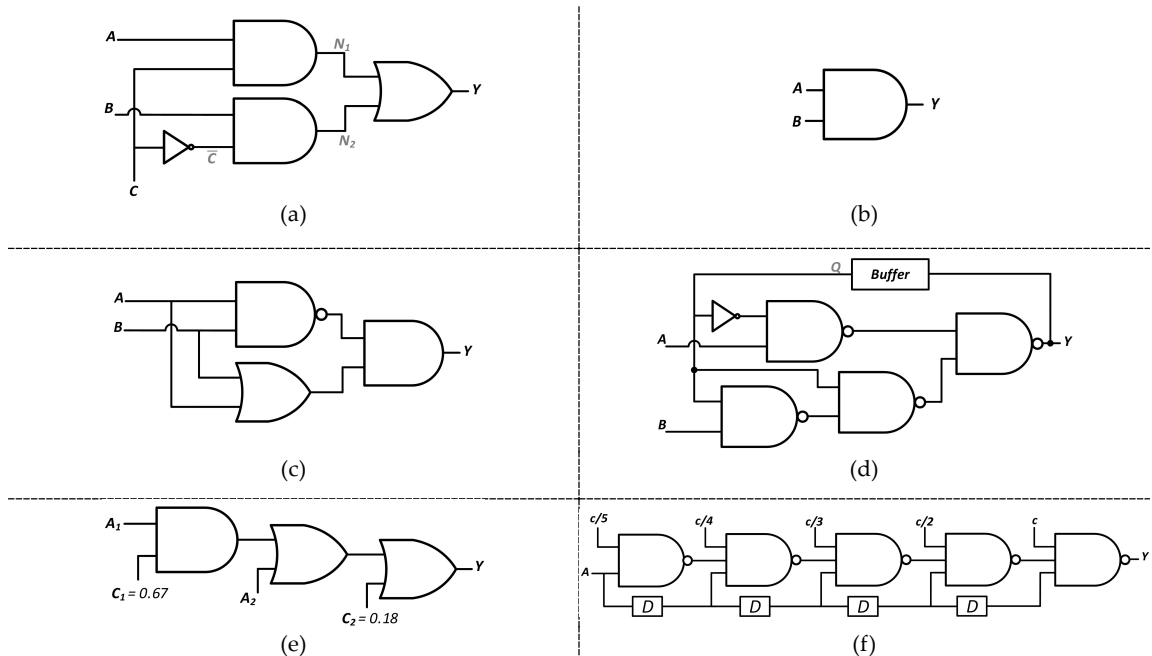

Fig. 5. Equivalent gate-level circuit of stochastic arithmetic operations a) scaled addition, b) multiplication, c) absolute value subtraction, d) scaled division, e) square root, and f) exponential based on supported primitive IMC logic gates.

In phase two, to implement the equivalent circuit of each stochastic operation in memory, we perform three steps: 1) presetting the memory cells, 2) writing input operands in memory cells (input initialization), and

3) performing logic steps. In the preset step, in addition to presetting the output cell according to the desired logic gate, the input cells need to be preset to the low resistance state to be ready for the stochastic bit generation. In the input initialization step, stochastic bit converting is done by utilizing the intrinsic stochasticity feature of STT-MRAMs, without needing additional cells or costly conventional SNG circuits. Finally, the logic step is similarly done as described in Section 2.2.

For clarification, Figure 6 illustrates the three-step implementation of the stochastic multiplication operation, equivalent to the AND gate, using 2T-1MTJ IMC method. In the preset step, input cells are preset to the low resistance state (logic '0'), and the output cell is preset to logic '1' according to the AND logic output cell preset value. In the second step, input operands are initialized, and the applied input writing pulses probabilistically switch the state of input MTJs to the high resistance, with probabilities depending on $V_I(A)$ and $V_I(B)$. Finally, in the third step, 2T-1MTJ cells operate in logic mode, by enabling logic transistors and applying the AND logic voltage (presented in [3,8]) to the *SL* lines of the input cells. For the implementation of other stochastic operations, a similar approach is taken.

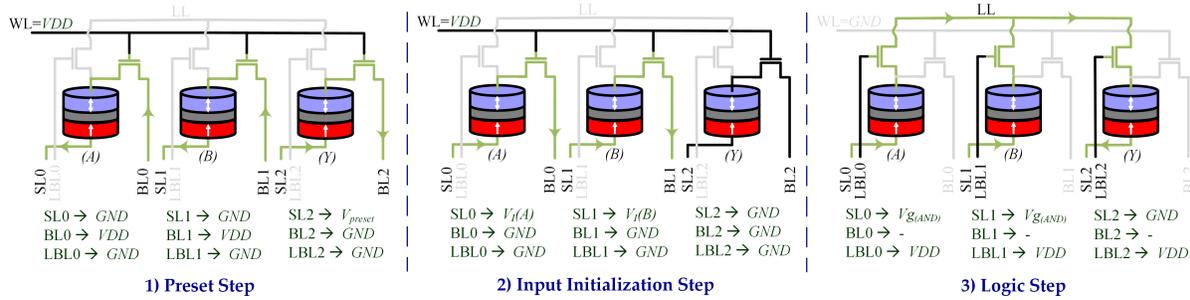

Fig. 6. Detailed three-step implementation of stochastic multiplication operation using 2T-1MTJ IMC method.

In Fig. 7(a) and Fig. 7(b), the sequence flow of a 4-bit in-memory addition operation in binary, and stochastic domains have been demonstrated, respectively. In this figure, the scheduled time and row/column number of logics have been shown. The binary addition has been performed based on the ripple carry adder structure by implementing four full adders (FAs) in memory. The effective equal logic function of FA for 2T-1MTJ IMC method is done through $C_{out} = NOT(MAJ_3(A,B,C))$ and $S = MAJ_5(A,B,C,C_{out},C_{out})$ [3,8]. In this regard, for the $FA_0$ in the first row, carry is obtained in the first step. Next, a copy is made by performing a BUFF logic, and ultimately, in the fifth step, the sum ($S_0$) is obtained. Also, the stochastic addition is performed based on the equivalent logic circuit described in Section 4.1 (Fig. 5(a)); the first four simultaneous NOT logics are performed on C followed by two-step AND logics, and finally, four simultaneous OR logics. It should be noted that due to the sequence flow similarity of the reset and input initialization steps for different operations, they are not demonstrated. Likewise, the sequence flow of other stochastic arithmetic operations is obtained.

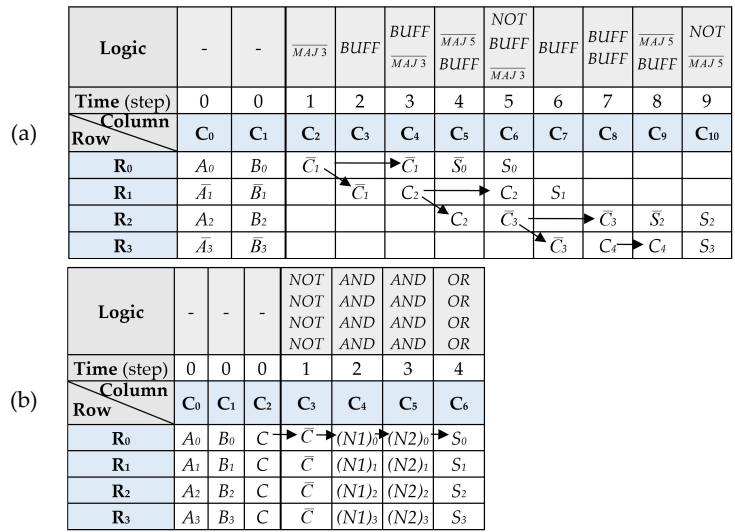

Fig. 7. Sequence flow of 4-bit in-memory addition operation in a) binary, and b) stochastic domains.

In this binary addition implementation, carry transfers require $2\times(n-1)$ cycles, and four cycles for odd bit-widths and three cycles for even bit-widths are needed to compute the most significant bit. Consequently, a 4-bit addition takes nine cycles to complete. Nevertheless, in stochastic implementation, regardless of the bitstream length, four cycles are taken. Therefore, this example clearly demonstrates the potential to effectively accelerate stochastic in-memory computations compared to binary in-memory computations by simplifying and parallelizing operations by 2T-1MTJ IMC method. Interestingly, this acceleration potential becomes more promising as the bit-with increases.

**4-2- Scheduling and Mapping**

It is possible to manually implement simple computations in memory by scheduling and mapping the constituent primitive logic gates. However, such a scheme is not practical for implementing the computation of large applications in memory. Therefore, an automatic solution is required. To do so, we present Algorithm 1, the first co-scheduling and mapping heuristic algorithm for 2T-1MTJ IMC method, which utilizes in-memory parallelization capability to reduce the total execution cycle time. In this way, we can effectively realize applications, such as neuromorphic ones in memory, within the SC domain. Algorithm 1 gets gate-level netlist of a given application and returns the execution cycle time of logic gates ($T$) and the mapped location of inputs and output operands of logic gates in 2T-1MTJ memory arrays. In this context, the maximum execution cycle time of logic gates shows the execution cycle time of the application.

In the algorithm, first, the netlist is sorted in topological order and the netlist depth is determined. This ensures that each gate is processed only after all its predecessors have been processed. The most common algorithm for this purpose is depth-first search (DFS) with a time complexity of $O(V + E)$, where $V$ is the number of vertices (gates) and $E$ is the number of edges (connections between gates) in the netlist. The memory array dimensions are initialized as $R_{available}$ and $C_{available}$. Then, the primary inputs (PIs) of the netlist are mapped. The algorithm maps the PIs with bit-width $q$ in a vertical layout to memory array columns. Each bit of the PI is assigned to consecutive rows in a column, and the column counter is incremented for the next PIs (In algorithm, *Memory(1:q, count)* denotes rows from 1 to $q$ and the column indexed by *count*). This initial mapping ensures that the PIs are suitably positioned for subsequent processing stages. Mapping the PIs requires iterating over each PI (lines 5-8), which has a time complexity of $O(p)$, where $p$ is the number of PIs.

Following the mapping of PIs, the algorithm iterates through each layer of the netlist. In this step, we aim to find subsets of the gates in each layer such that the gates in each subset can be parallelized. Based on 2T-1MTJ IMC method, three constraints must be met for gates to be parallelized: 1) the gates must be of the same type, 2) the gates must not have same input, and 3) the gates must be input-column-aligned. Therefore, we create subsets to partition the gates based on these constraints. For each layer, in line 11, subsets of gates are created based on identical gate types and the absence of connections (common fan-in). These subsets are then sorted in descending order based on the average of their inverse topological order values (lines 12-13). The inverse topological order presents the distance of gate to primary output. This sorting prioritizes gates that should be executed earlier, improving the execution flow. The time complexity of creating subsets and sorting are $O(n^2)$ and $O(n \times logn)$, respectively, where $n$ is the number of gates in the netlist.

Next, within each subset, the algorithm processes gates to ensure their inputs are in the same row. It is a constraint necessary for performing logic based on the IMC method. In this regard, for gates requiring two inputs, the algorithm checks if the inputs are in different rows and, if so, copies the second input to the next available column in the same row as the first input (lines 15-22).

Afterward, we partition gates based on the third logic parallelization constraint, input-column-alignment. Thus, the algorithm creates new subsets based on the alignment of input columns within each subset (line 23). It then iterates through these subsets, mapping each gate's output to the next available column in the same row as its input. This mapping is done while incrementing a cycle counter to track the number of execution cycles (lines 24-30).

The algorithm returns the execution cycles and the row-column coordinates for each gate in memory array. This algorithm ensures that the gates are scheduled and mapped correctly by considering 2T-1MTJ IMC method constraints. Also, enabling intra-subarray parallelization and performing possible gates in a parallel manner reduces execution cycle time and enhances the overall performance of Stoch-IMC. Since the number of required cycles for input initialization can vary based on the utilized in-memory binary to stochastic (BtoS) conversion approach, these cycles have been ignored here, but are later added to the total execution cycle time. Additionally, Algorithm 1 supports one- and two-input gates, but it can be extended to support gates with more inputs if necessary.

In conclusion, although the algorithm consists of nested loops, each gate is visited only once, except during the creation of subsets. The overall timing complexity of the algorithm is determined by the creation of subsets' part as O($n^2$). Importantly, because the in-memory scheduling and mapping algorithm is executed only once during the design phase for each application, this time complexity is not a major issue.

It should be noted that if the required memory array size for computing the equivalent circuit of the application exceeds the memory subarray size, the circuit must be partitioned into smaller parts before being processed by Algorithm 1. The algorithm runs on these partitioned circuits sequentially, completing each partition before moving to the next. Consequently, the index term $q$ in the input initialization part of the algorithm (lines 5-8) represents the sub-bitstream length of the PIs of the partitioned input circuit. For example, if an application's equivalent stochastic circuit can be processed without partitioning, $q$ equals the full bitstream length. Conversely, if only the stochastic circuit for one bit of the application can be processed, $q=1$ in Algorithm 1.

---

**Algorithm 1:** Stoch-IMC scheduling and mapping in memory array

**Input:** Netlist of supported gates $G = [g_1, g_2, …, g_n]$
**Output:** Execution cycle time, and mapping location of gates

1. $G_{sorted}$ = topological_order_sort(G);
2. $L$ = depth of netlist;
3. Initialize $R_{available}$ and $C_{available}$;
4. count = 1;
5. **foreach** $PI_{i,[1:q]}$ in PI **do**
6.    Map $PI_{i,[1:q]}$ to Memory(1:q, count);
7.    count++;
8. **end foreach**
9. cycle = 0;
10. **for** i=1 to L **do**
11.    Create subsets (Subset) of identical gate type and not having input connections from gates in $G_{sorted}(i)$;
12.    Compute the average of inverse topological order value of gates in each subset;
13.    Sort subsets in descent order;
14.    **foreach** $s \in$ Subset **do**
15.       **foreach** g in s **do**
16.          **if** gate_type(g) = 2 **then**
17.             **if** (inputs of g are not in the same row) **then**
18.                Copy the second input of g to next available column in the same row as the first input;
19.                cycle++;
20.             **end if**
21.          **end if**
22.       **end foreach**
23.    Create subsets (SubsetA) of input-column-aligned gates from s;
24.    **foreach** sA $\in$ SubsetA **do**
25.       cycle++;
26.       **foreach** g in sA **do**
27.          Map output of g to the next available column in the same row as the input of g;
28.          T(g) = cycle;
29.       **end foreach**
30.    **end foreach**
31.    **end foreach**
32. **end for**
33. **return** (T, row-column number of I/O operands of gates);

---

### 4-3- Stoch-IMC Memory Architecture

In the previous Section, we presented an in-memory scheduling and mapping algorithm that works on a memory (sub)array. Nonetheless, in general, under any size of subarray (SA) and any number of SAs in each bank, there may be a lack of in-memory computation space when different applications are considered. For instance, the computation of one bit of the bitstream of an application may need more cell space to perform as the memory array size. However, there are limitations on the size of SAs and the number of them in each bank of memory, so we cannot consider very large SA sizes [40]. It is due to in-memory computation reliability (high $I \times R$ drop and noise vulnerability), memory optimality design, and physical limitations.

As a result, there is a need to provide a novel memory architecture to reach high-performance in-memory stochastic computation for large applications while enabling the possibility of parallelizing bit computations. The Stoch-IMC memory architecture, which is shown in Fig. 8 comprises multiple banks, each of which includes multiple groups of subarrays (*n*) and each group consists of multiple local subarrays (*m*) called [*n*, *m*] configuration. The subarrays of each group are connected by the local bus, and the groups are connected together by the global bus. Each group of subarrays has a (1-bit input with $\lfloor \log(m) \rfloor$+1 bit register) local accumulator unit that counts the number of ones of the in-memory stochastic computation result. Also, there is a ($\lfloor \log(m) \rfloor$+1 bit input with $\lfloor \log(n \times m) \rfloor$+1 bit register) global accumulator that provides the result of the computation in the binary domain by summing up the total number of ones of the partial results. Note that by understanding the number of ones in the bitstream, we have the binary value of the result. There are also bank I/O, bank controller, and $2^{resolution}$ Byte binary to stochastic (BtoS) memory that determine the corresponding voltage pulse equal to the binary input. For instance, for 8-bit binary and 256-bit bitstream resolution, the BtoS memory size is equal to 256B. Therefore, by exploiting the intrinsic stochasticity feature of memristors and applying the corresponding voltage pulse, one can generate stochastic numbers without the need for costly CMOS-based random number generators.

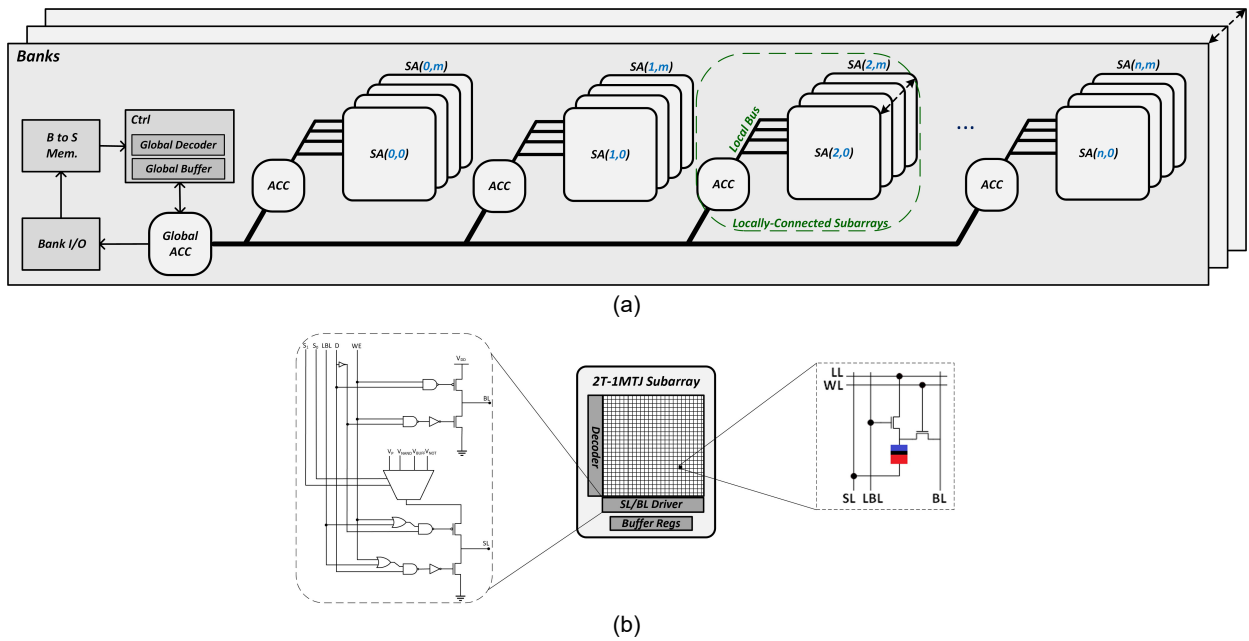

Fig. 8. Overview of Stoch-IMC 2T-1MTJ-based [*n*, *m*] configuration memory architecture, a) bank, and b) subarray.

Subarrays as the in-memory processing elements are arranged in a regular square form. Therefore, if we have *m* locally connected SAs in each group and *n* number of groups, *n*=*m*. In the bit-parallel Stoch-IMC memory architecture, by utilizing independent computation of bits in SC, the computation of the bits of the bitstream is performed individually in different subarrays. In this regard, first, computations are assigned to the subarrays of one group. Then, if the application bitstream length is greater than *m* (number of SA in each group), the rest of the computations are assigned to the next group, and so on to assign all bits of the bitstream. Finally, if the bitstream length is greater than *n*×*m*, we can either pipeline the data based on reusing one bank, which minimizes the area at the expense of increased latency, or parallelize it over many banks, which processes the data with decreased latency. In the case of pipelining, each pair of sub-bitstreams is processed, the output is stored, and then another pair of sub-bitstreams is processed. If the bitstream is divided into *K* sub-bitstreams, the number of processing cycles required to handle the whole computation is improved by a factor of *K* in the parallelizing approach compared to pipelining. However, a few more cycles are also required for data transfers. It is also feasible to combine the parallel and pipeline approaches to achieve a better area and latency trade-off, but this incurs more complicated implementation and partition management.

The provided square and local base structure architecture reduces the intermediate data transfers and their cost in the result accumulation phase in contrast with the ungrouped globally connected structure. For instance, assume the application bitstream length is 256 bits and *n*, *m* = 16. The accumulation phase consumes 32 steps, it takes 16 steps for local accumulation in groups and 16 steps to accumulate the results of the different groups while 256 steps are required in an ungrouped globally connected structure. So, in general, it takes $n+m$ steps instead of $n \times m$ steps.

## 5- Evaluation

### 5-1- Evaluation Setup and Flow

We conduct SPICE simulations to evaluate the operations in memory at the circuit level, based on predictive technology model (PTM) of CMOS [41] and MTJ model [8,42]. The specific parameters of the MTJ model are provided in Table 1. We extract circuit-level parameters of the memory array and its peripheral circuitry by using NVSim. In application-level evaluations, the Stoch-IMC memory architecture includes $n$=16 groups with $m$=16 subarrays of size 256×256. The subarray size has been determined based on in-memory computation reliability considerations, including noise margin and $I \times R$ voltage drop. Although we can consider several banks in the memory architecture, to make a fair comparison with the related work [22], we consider only one bank. Also, we proceed with a pipeline approach if the bitstream length is greater than $n \times m$. Finally, we use MATLAB to evaluate the applications' output accuracy.

TABLE 1
THE PHYSICAL PARAMETERS OF MTJ ELEMENT

| Parameter | Description | Value |
|---|---|---|
| Type | Perpendicular magnetic anisotropy MTJ | PMA_MTJ |
| a, b | Length and width of surface | 10 nm |
| $R_P$, $R_{AP}$ | Low/High resistance | 12.7, 76.3 KΩ |
| TMR | Tunneling magnetoresistance ratio | 500% |
| $J_c$ | Critical switching current density | $10^6$ A/cm$^2$ |
| $I_c$ | Critical switching current | 0.79μA |
| $T_{switching}$ | Time of MTJ switching | 1ns |

In our assessment, we evaluate different design parameters such as performance, energy consumption, area, and accuracy. The total time cycle is considered as the performance metric, and regarding the area the number of used memory cells is defined. Also, the accuracy is obtained through application-level analysis in terms of the output accuracy. The energy consumption is calculated as follows:

$$E_{total} = BL \times E_{computation} + E_{peripheral} \quad (3)$$

where $BL$ is the bitstream length of the stochastic number and $E_{computation}$ and $E_{peripheral}$ are, respectively, the consumed in-memory computations energy and the peripheral circuitry energy. Based on the in-memory implementation of the stochastic arithmetic operations and the energy consumption of each logic gate, $E_{computation}$ is calculated as,

$$E_{computation} = N_{preset} E_{preset} + N_{SBG} E_{SBG} + \sum N_g E_g \quad (4)$$

where $N_{preset}$ and $N_{SBG}$ are the number of presets and stochastic bit generations (SBGs), and their energy consumption has been denoted by $E_{preset}$ and $E_{SBG}$, respectively. It is worth mentioning that the combination of $V_p$ and $t_P$ that leads to the lowest switching energy for the stochastic bit generation has been considered in our evaluations. It can be obtained for the desired switching probability (e.g., $P_{sw}$ = 0.5) through (1), (2), and $E=V_p^2 \times t_P/R_{MTJ}$ equation [33]. Finally, the third term of (4), where g ∈ {NOT, BUFF, NAND, NOR, $\overline{MAJ3}$, $\overline{MAJ5}$}, represents the consumed energy of used logic gates. We conducted SPICE simulations to determine the energy consumption of each IMC-supported logic gate. The energy values for the six logic gates and the PRESET operation are as follows: NOT=30.7aJ, BUFF=73.8aJ, NAND=28.7aJ, NOR=8.4aJ, $\overline{MAJ3}$=7.6aJ, $\overline{MAJ5}$=6.3aJ, and PRESET=26.1aJ. These values are multiplied by the number of each gate to calculate the total energy. Consequently, based on (4), the evaluation included necessary operations for input, intermediate, and output cells, as well as intra-subarray data transfers.

The $E_{peripheral}$ comprises the subarray peripheral circuitry energy (including the modified SA driver), BtoS memory energy, and accumulations' consumed energy for stochastic to binary (StoB) conversion. In general, for [$n$, $m$] configuration of Stoch-IMC architecture with ⌈log($n \times m$)⌉ bit computation resolution, $n \times m$ local accumulations (1-bit input with ⌈log($m$)⌉+1 bit register) and $n$ global accumulations (⌈log($m$)⌉+1 bit input with ⌈log($n \times m$)⌉+1 bit register) are done. The subarray peripheral circuitry and BtoS memory energies are obtained by NVSim, and the accumulations' energy is obtained by multiplying the number of them to their (local/global accumulator) extracted energy from implementation by 15$nm$ Nangate standard cell library in Synopsys Design Compiler.

While 2T-1MTJ IMC method is a reliable IMC architecture, we enhance the reliability of computations in Stoch-IMC by leveraging a subset of supported logic gates with maximum computation reliability, including NOT, BUFF, and NAND [3,8]. In our evaluations, 8-bit resolution has been considered. So, binary in-memory computations have been done based on 8-bit fixed-point numbers, and stochastic ones are done on 256-bit bitstream length. We consider ripple carry addition and Wallace tree multiplication for implementing an 8-bit binary adder and multiplier in memory. Also, based on full-subtraction and non-storing array division units, the 8-bit binary subtract and divide arithmetic operations have been implemented in memory. Eventually, the binary square root and exponential are implemented by utilizing three steps of the Newton-Raphson method and the Maclaurin expansion method (fifth-order polynomial), respectively [20,38].

### 5-2- Arithmetic Operations Evaluation

Theoretically, any function that can be executed in CMOS-based SC can also be implemented in Stoch-IMC. However, this study specifically focuses on the arithmetic operations required for evaluating applications. Based on the aforesaid evaluation flow, we evaluated the minimum array size, area (based on the number of used memory cells), total time steps (cycles), and energy consumption of the arithmetic operations for Stoch-IMC, a well-known digital stochastic IMC work [22], and the binary implementation of these operations in memory based on 2T-1MTJ IMC method (Binary IMC) [3,8]. The results of the computation part have been summarized in Table 2. In this regard, the area, time steps, and energy results have been normalized to the corresponding ones of the binary in-memory implementation of each arithmetic operation.

TABLE 2
COMPARISON OF STOCH-IMC TO RELATED WORK IN DIFFERENT ARITHMETIC OPERATIONS (NORM. TO IN-MEMORY BINARY IMPLEMENTATION).

| *Arithmetic Operation* | *Minimum Array Size (Row×Col.)* | | | *Area (number of used cells)* | | *Total Time steps* | | *Energy Consumption* |
| --- | --- | --- | --- | --- | --- | --- | --- | --- |
| | *Binary IMC* | *Stochastic IMC* | | *Stochastic IMC* | | *Stochastic IMC* | | *Stochastic IMC* |
| | | [22] | *This work* | [22] | *This work* | [22] | *This work* | *This work* |
| Scaled Addition | *1×88* | *1×7* | *256×7* | *0.080X* | *20.36X* | *14.3X* | *0.056X* | *14.640X* |
| Multiplication | *16×161* | *1×4* | *256×4* | *0.002X* | *0.397X* | *5.1X* | *0.012X* | *0.983X* |
| Absolute Value Subtraction | *1×90* | *1×8* | *256×8* | *0.090X* | *22.75X* | *22.5X* | *0.088X* | *15.379X* |
| Scaled Division | *8×130* | *1×13* | *256×13* | *0.013X* | *3.2X* | *2.0X* | *0.008X* | *2.116X* |
| Square Root | *32×1413* | *1×10* | *256×10* | *0.0002X* | *0.056X* | *0.49X* | *0.002X* | *0.253X* |
| Exponential | *17×1255* | *1×31* | *256×31* | *0.001X* | *0.372X* | *4.86X* | *0.019X* | *0.857X* |

The results indicate that, across all arithmetic operations, this work achieves a significant reduction in total time steps compared to both binary IMC implementation and [22]. This improvement is primarily attributed to 1) enabling in-memory intra-subarray bit parallelization, and 2) utilizing low-complexity stochastic arithmetic operation implementations. However, the timing results of [22] have deteriorated in most arithmetic operations compared to the binary implementation due to the bit-serial computation behavior on long bitstream.

In terms of array size and area, this work achieves better results in some arithmetic operations compared to the binary IMC while requiring more cell usage in others. This outcome arises from the trade-off between the decreased complexity of the equivalent stochastic circuit of the operations and the increased bit-width in stochastic computation in comparison to binary computation. For example, the results for scaled addition show higher area overheads than the binary implementation because 8-bit binary addition is not a particularly complex arithmetic operation compared to its stochastic equivalent. However, the results demonstrate the considerably lower complexity of the stochastic square root implementation versus its binary counterpart.

Regarding cell usage, two points should be noted: 1) although our work shows the increased area (cell usage) in some arithmetic operations compared to the binary IMC, the area of memristive cells is smaller than that of conventional memory cells [26], making the imposed area overhead in some operations still acceptable. 2) The minimum required array size for binary IMC implementation of some operations is large enough to encounter reliability challenges (i.e., $I \times R$ voltage drop and low noise margin) based on 2T-1MTJ IMC method [40].

In terms of energy, our results are better in some functions, while in others, the binary IMC implementation performs better. Although the equivalent circuit for stochastic operations is simpler than that for binary implementations, the repeated computation based on the bitstream length can lead to higher energy consumption in some cases. We have used a 256-bit bitstream equivalent to an 8-bit binary resolution to keep the computation error negligible. However, it is possible to choose a shorter bitstream length to create a suitable trade-off among design parameters in approximable applications, which can lead to energy improvement. It should be noted that research [22] did not report their arithmetic operations' energy results. Nonetheless, due to the similar stochastic implementation of each bit of arithmetic operations with this work, the results may be in the same order.

In conclusion, the findings demonstrate that Stoch-IMC can be an effective solution for a wide range of applications using various arithmetic operations. Specifically, it excels in applications involving many multiplication, scaled division, square root, and exponential operations. Additionally, Stoch-IMC is efficient for applications that require high-order polynomials and Maclaurin expansions of complex operations.

**5-3- Applications Evaluation**

*5.3.1. Applications Definition*

To evaluate the merit of Stoch-IMC, it is essential to investigate it in different applications. In this regard, we showcase the effectiveness of Stoch-IMC in neuromorphic and image processing applications, including local image thresholding [38], Bayesian inference for object location [36], Bayesian belief network for predicting heart disasters [36], and kernel density estimation [37].

In local image thresholding (LIT) application, a ($n \times n$ pixels) window is applied within a subsection of a degraded input image. For each subsection, a threshold ($T$) is determined by (5), where $A$ denotes the pixel intensity at a given point. In this context, the mean ($\bar{A}$) and standard deviation ($\sigma A$) of all the pixels within the window are computed, where $\sigma A$ is determined using (6). The effective gate-level SC realization of this application has been shown in Fig. 9(a). In this figure, although $A_1$ and $A_2$ have the same value, they were generated separately.

$$T(x,y) = \overline{A(x,y)} \times \left( \frac{\sigma A(x,y)+1}{2} \right) \tag{5}$$

$$\sigma A(x,y) = \sqrt{\left| \overline{A(x,y)^2} - (\overline{A(x,y)})^2 \right|} \tag{6}$$

For object location (OL) application, based on gathered distance ($D$) and bearing ($B$) data from three sensors a Bayesian inference system is introduced. In this regard, the object location probability ($p(x,y)$) is calculated by (7) based on conditional probabilities in the Bayesian inference system. Figure 9(b) illustrates the effective gate-level SC implementation of this application.

$$p(x,y) = \prod_j p(B_j|x,y) \times p(D_j|x,y) \tag{7}$$

The third application is a Bayesian belief network for heart disaster prediction (HDP). It uses a directed acyclic graph to represent random variables and their conditional dependencies. In this regard, the heart disaster probability ($P(HD)$) is calculated by (8), where $P(HD|E,D)$ indicates the probability considering solely exercise ($E$) and diet ($D$). Additionally, $P(BP)$ and $P(CP)$ denote the high blood pressure and chest pain probabilities, respectively. Also, $P(HD|E,D)$ is calculated using (9), which incorporates $P(D)$, $P(E)$, and $P(E,D)$ to represent the probabilities for a good diet, regular exercise, and both combined. Figure 9(c) depicts the effective gate-level circuit used to compute (8).

$$P(HD) = \frac{P(BP) \times P(CP) \times P(HD|E,D)}{P(BP) \times P(CP) \times P(HD|E,D) + P(\overline{BP}) \times P(\overline{CP}) \times P(\overline{HD}|E,D)} \tag{8}$$

$$P(HD|E,D) = [P(E,D) \times P(D) + P(E,\overline{D}) \times P(\overline{D})] \times P(E) + [P(\overline{E},D) \times P(D) + P(\overline{E},\overline{D}) \times P(\overline{D})] \times P(\overline{E}) \tag{9}$$

Finally, the last examined application is Kernel Density Estimation (KDE), which relies on continuously updating recent information. Regarding this, the pixel intensity values ($X$), derived from recent history ($X_t$, $X_{t-1}$, …, $X_{t-N}$), are utilized to calculate the probability density function (PDF), as outlined in (10). The corresponding effective gate-level stochastic circuit is illustrated in Fig. 9(d).

$$PDF(X_t) = \frac{1}{N} \sum_{i=1}^{N} e^{-4|X_t - X_{t-i}|} \tag{10}$$

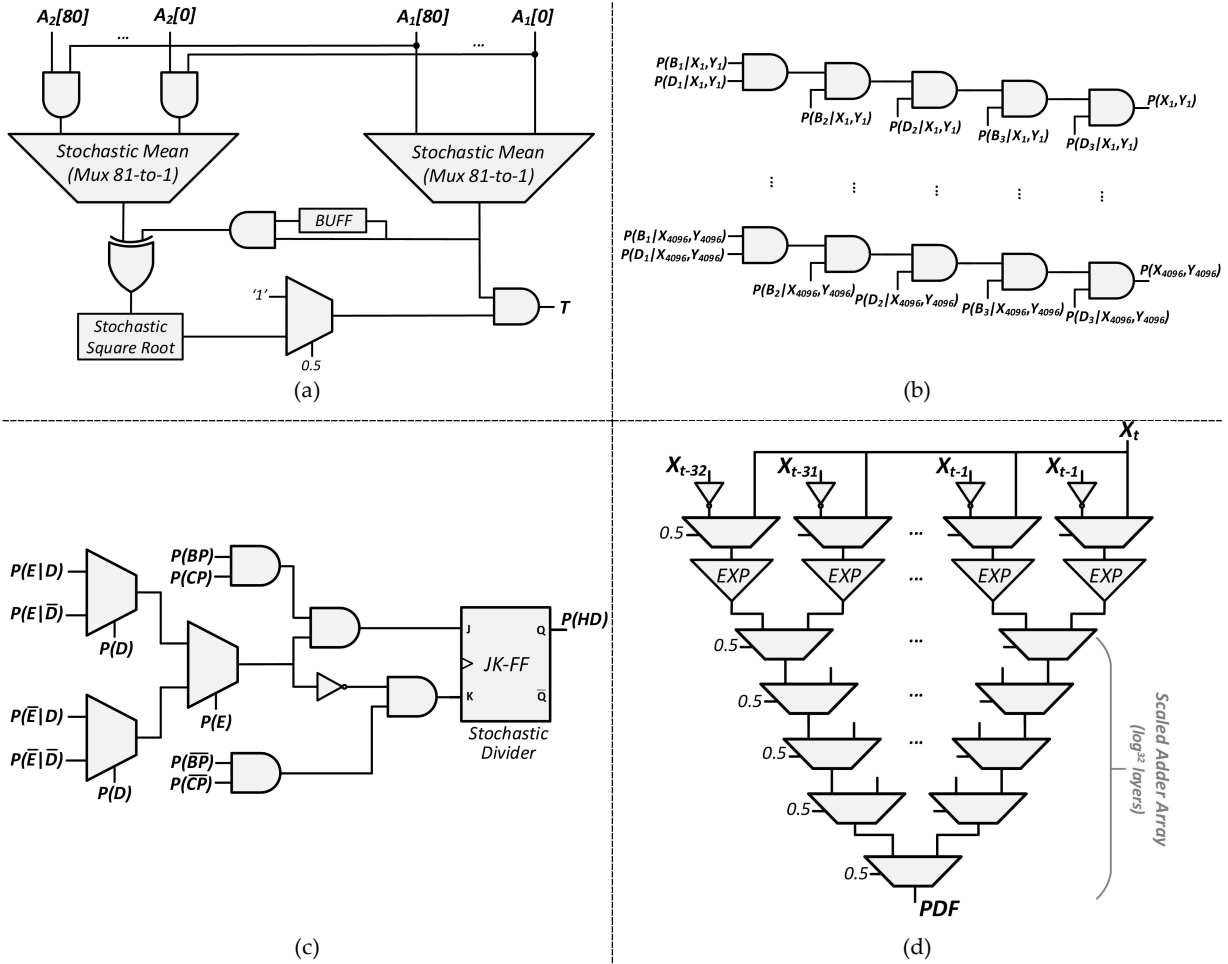

Fig. 9. Effective stochastic gate-level circuit of applications: a) local image thresholding [38], b) object location [36], c) heart disaster prediction [36], and d) kernel density estimation [37].

### 5.3.2. Evaluation Results

We demonstrate the efficiency of Stoch-IMC in terms of total time steps, energy consumption, and cell usage in different applications, and the results are shown in Table 3. The results of this work and the related work [22] have been normalized to equivalent in-memory binary implementation. In time step evaluations, the delay overhead of preset output cell of gates (except the last one) overlaps with consecutive logic operations. Also, it should be mentioned that the total time steps and energy consumption results of [22] have been replicated due to differences in the baseline implementation as well as the utilized memory and transistor models. In this work, the baseline binary IMC relies on effective intra-subarray parallelization-enabled implementation [3,8], and the utilized model parameters are shown in Table 1.

In this study, the stochastic circuit of each application has been provided as input to Algorithm 1, and the scheduling and mapping of the circuit in memory is achieved. The local image thresholding and object location applications have been evaluated on a 9×9 window size and a 64×64 2D grid, respectively. To implement the $e^{-4x}$ operation in the stochastic circuit of kernel density estimation application, $e^{-4/5x}$ was first estimated using the fifth order of Maclaurin expansion to keep data in the range of unipolar encoding. This was then achieved through five stages of $e^{-4/5x}$ multiplication.

The results indicate that Stoch-IMC significantly improves total time steps across all four applications, achieving average improvement of 135.7X and 124.2X (geometrical mean) compared to binary IMC and [22], respectively. The main reasons behind this performance improvement are: 1) the simplicity of stochastic computation circuits, 2) the proposed bit-parallel computing architecture, and 3) the intra-subarray parallelization-enabled scheduling and mapping of Algorithm 1. It is important to note that effectively partitioning the input stochastic circuit before giving it to Algorithm 1 can yield further performance improvement. For instance, in the objection location application (Fig. 9(b)), the stochastic circuit has been partitioned into the computation of each pixel of the 2D grid, and it has been given as input to Algorithm 1 ($p$=6 and $q$=1). Thus, employing [16, 16] configuration of Stoch-IMC architecture allows for parallel computation of all 256 bits of the bitstream, necessitating 4096 stages to complete the entire pixel computation of the 2D grid. In contrast, by partitioning into batches of 16 pixels of stochastic circuits and providing this as input to the algorithm, only 256 stages are required. As another example, for kernel density estimation application, $p$=1 and $q$=33 were considered in Algorithm 1 as an effective partitioning strategy.

TABLE 3
COMPARISON OF STOCH-IMC TO RELATED WORK IN DIFFERENT APPLICATIONS (NORM. TO IN-MEMORY BINARY IMPLEMENTATION).

| Application | Subarray Size (Row×Col.) | | | Area | | Total Time steps | | Energy Consumption | |
|---|---|---|---|---|---|---|---|---|---|
| | Binary-IMC | [22] | This work | [22] | This work | [22] | This work | [22] | This work |
| Local Image Thresholding | 2048×4096 | 128×128 | 128×128 | 0.048X | 12.49X | 0.463X | 0.003X | 5.694X | 5.711X |
| Object Location | 16×1024 | 1×16 | 1×16 | 0.005X | 1.31X | 5.908X | 0.085X | 0.816X | 1.244X |
| Heart Disaster Prediction | 128×1024 | 8×32 | 8×32 | 0.005X | 1.31X | 0.454X | 0.004X | 0.046X | 0.056X |
| Kernel Density Estimation | 512×2048 | 32×64 | 32×64 | 0.022X | 5.72X | 0.565X | 0.003X | 0.449X | 0.455X |

The energy consumption results show a 1.5X reduction on average compared to binary IMC. In this regard, the findings show decreased energy consumption for heart disaster prediction and kernel density estimation applications, while an increase was observed for the local image thresholding and object location applications. Among the four evaluated applications, the local image thresholding application consumed 5.711X more energy than the binary IMC implementation. It is attributed to 1) the energy trade-off between decreased complexity of the equivalent stochastic circuit and increased bit-width in stochastic versus binary computations, 2) the number of input values of the application, which require two steps in stochastic input initialization (preset and stochastic writing) as opposed to binary input initialization, and 3) the need for numerous copy operations (by BUFF logic) to enable intra-subarray parallelism based on Algorithm 1 for this application. Additionally, compared to the stochastic method [22], a limited energy overhead has been imposed. This overhead primarily comes from the StoB conversion mechanism based on the accumulation structure presented in the Stoch-IMC. Nonetheless, this energy overhead is considered acceptable due to the significant reduction in energy consumption of IMC compared to non-IMC systems [8].

Our results are derived from the [$n$, $m$] configuration of the presented architecture, in which the number of utilized cells in $n$×$m$ subarrays multiplied by cell area has been employed. This multiple-subarray configuration, accounting for the imposed overhead of the added peripheral circuits (including accumulators and BtoS memory), consumes more area compared to [22], which employs a single subarray. Nonetheless, it is important to note that due to memristors having a smaller footprint than CMOS-based memories [26], Stoch-IMC remains competitive in terms of area when compared to stochastic methods based on conventional charge-based memories. Furthermore, based on the subarray size results, this work requires the same subarray size as [22] across all four applications, while necessitating significantly fewer rows and columns than the binary IMC implementation. It is an important parameter in the reliability of 2T-1MTJ IMC method due to its susceptibility to the parasitic effect of $I$×$R$ voltage drop [3,22,40]. The minimal required subarray size for applications in Stoch-IMC is within the reliable range of 2T-1MTJ in-memory computations [3,22,40]. Therefore, although Stoch-IMC may consume more energy or area than binary IMC in some applications, it becomes a preferable method when reliable computation is essential.

In conclusion, limitations of the presented scheme can be summarized as: 1) Achieving high computational accuracy requires large memory cells to represent numbers precisely in stochastic computing data representation, as compared to compact binary computation. Therefore, for applications that cannot be approximated, the memory size would be large. 2) While enabling bit parallelization of applications leads to significant performance improvements, trade-off between complexity of the equivalent stochastic circuit of the application and bit-width in stochastic computations can limit energy improvement. 3) The presented parallel computation scheme operates on the multiple-subarray architecture. This multiple-subarray architecture, which accounts for the imposed overhead of added peripheral circuits (including accumulators and BtoS memory), consumes more area than schemes that employ only one single subarray [22].

**Energy breakdown:** The energy breakdown evaluation across four applications has been performed for binary IMC, method [22], and Stoch-IMC and the results are demonstrated in Fig. 10. This figure shows how energy is distributed among logic, reset, input initialization, and peripheral circuitry across applications for each method. The energy distribution indicates that the main areas of energy usage are the logic and reset steps in all three methods and applications. The percentage of energy consumption for logic operations is generally lower in stochastic-based methods, including [22] and Stoch-IMC, compared to binary IMC. This reduction can be attributed to the less complex circuits used in stochastic approach. Conversely, the percentage of energy consumption for reset operations tends to be higher in stochastic-based methods than in binary IMC. This increase is due to the necessity of performing reset operations prior to both input initialization and logic steps (for determining the logic operation) in stochastic-based methods.

The contribution of input initialization energy of binary IMC method (through deterministic write operation) is less than that in stochastic-based methods (through stochastic write operation) across all four applications. This difference primarily arises from the greater number of write operations required in stochastic data representation (256-bit) compared to binary representation (8-bit). However, the energy consumed by input initialization and peripheral circuitry constitutes a minority of the total energy. In the case of binary IMC and method [22], the peripheral circuits include the peripheral subarray circuitry (such as SL/BL drivers). In contrast, Stoch-IMC's peripheral circuitry encompasses not only the peripheral subarray circuitry but also accumulators and BtoS memory, resulting in higher peripheral circuitry power consumption than both binary IMC and method [22]. Additionally, it is noteworthy that a comparison of power usage for different subarray sizes indicated that power consumption of peripheral circuitry increases with larger subarray sizes.

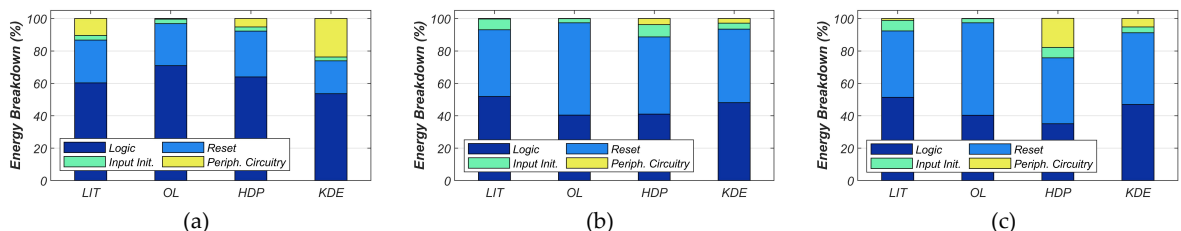

Fig. 10. Energy breakdown of a) binary IMC, b) method [22], and c) Stoch-IMC for different applications.

**Lifetime:** The lifetime of non-volatile memories is an essential factor for reliability. It is directly proportional to endurance ($E_{Max}$) and memory capacity ($C$) while being inversely proportional to the memory access traffic ($B$), which varies depending on application [43],

$$Lifetime \propto \frac{E_{Max} \times C}{B} \qquad (11).$$

Since endurance is a constant value dependent on memory technology (e.g., >$10^{15}$ for STT-MRAM), the memory capacity and access rate determine the lifetime improvement when comparing different methods that utilize the same memory technology. To evaluate the lifetime of Stoch-IMC, we can compare it in terms of memory access rate (specifically, write access, as it is the dominant factor affecting endurance) and memory capacity with other methods using identical memory technology across different applications. Therefore, we can calculate the product of the inverse ratio of the memory write access rates and the direct ratio of the subarray size and number of subarrays, making this product proportional to the lifetime improvement.

Equation (11) estimates the lifetime, assuming that memory accesses are uniformly distributed across all memory cells. However, since none of the evaluating methods employ such a mechanism, we use the number of utilized cells instead of the total memory size to achieve a more precise evaluation. The normalized lifetime improvement results of Stoch-IMC and method [22] compared to binary IMC are presented in Fig. 11. The results indicate that Stoch-IMC, by achieving a better trade-off between the number of used cells and memory accesses, achieves on average lifetime improvement of 4.9X and 216.3X over binary IMC and method [22] in various applications, respectively. The lifetime deficiency of the method [22] is mainly attributed to its bit-serial computation approach within a single subarray, which causes certain memory cells to experience more access stress. In contrast, Stoch-IMC distributes bit computation across multiple subarrays.

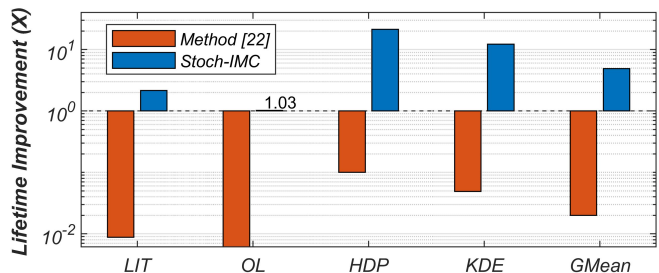

Fig. 11. The lifetime improvement of method [22], and Stoch-IMC for different applications.

**Bitflip:** Soft error is one of the threats to reliability in digital systems that manifests itself as a data bitflip in digital systems. Moreover, STT-MRAMs' errors during reading, writing, and computation—primarily caused by the stochastic switching behavior of the MTJ in the STT-MRAM cell—can result in bitflips. Therefore, minimizing the negative impact of bitflip is critical in STT-MRAM-based IMC systems. In conventional binary IMC methods, if a bitflip occurs in the most significant bits, a substantial output error occurs. However, stochastic-based IMC methods offer a clear advantage since all bits in the stochastic data representation hold equal importance, thereby minimizing the negative effects of bitflips on the overall data value.

To demonstrate the superiority of Stoch-IMC over binary IMC, the average output error of different applications, when various degrees of fault are injected, is illustrated in Table 4. In this context, fault injection (manifested as bitflips) has been randomly applied to the input/output nodes of the stochastic arithmetic operations in applications. The results indicate that when the injected bitflip rate exceeds 5%, Stoch-IMC outperforms binary IMC. Conversely, at bitflip rates of less than 5%, the output error of Stoch-IMC is greater than that of binary IMC. However, this error is primarily due to stochastic approximation errors rather than those from fault injection. Furthermore, Stoch-IMC achieved an error rate below 6.5% in all four applications, even at a high injected bitflip rate of 20%. Therefore, in environments prone to soft errors or where the data can be noisy and erroneous, Stoch-IMC could be beneficial.

TABLE 4
THE AVERAGE OUTPUT ERROR (%) OF BINARY-IMC (8-BIT) AND STOCH-IMC (256-BIT) IN DIFFERENT APPLICATIONS WHEN BIT FLIP RATE CHANGES.

| Method | | *Binary-IMC* | | | | | *Stoch-IMC* | | | | |
|---|---|---|---|---|---|---|---|---|---|---|---|
| *Injected Bitflip Rate (%)* | | *0* | *5* | *10* | *15* | *20* | *0* | *5* | *10* | *15* | *20* |
| *Application* | Local Image Thresholding | 0 | 7.9 | 32 | 35 | 40 | 0.9 | 2.4 | 4.2 | 5.5 | 6.4 |
| | Object Location | 0 | 2.3 | 3.5 | 4.6 | 16.8 | 0.06 | 0.08 | 0.09 | 0.15 | 0.18 |
| | Heart Disaster Prediction | 0 | 1.2 | 2.2 | 3.4 | 13.7 | 0.03 | 0.05 | 0.07 | 0.10 | 0.13 |
| | Kernel Density Estimation | 0 | 5.6 | 10.1 | 14.2 | 18.3 | 1.20 | 1.36 | 1.39 | 1.49 | 1.53 |

## 6- Conclusion

In this paper, we proposed a novel STT-MRAM-based stochastic in-memory computing architecture called Stoch-IMC which leverages bit-parallel stochastic computations in memory along with the presented in-memory scheduling and mapping algorithm. We integrated key advantages of SC and IMC, such as low computation complexity and high bit-parallel computation capability, to present effective in-memory computations. Our comprehensive evaluation across various abstraction levels and stochastic applications, compared to binary IMC and in-memory SC methods, demonstrated on average 135.7X and 124.2X performance improvements, respectively. Additionally, energy consumption comparisons with binary IMC and in-memory SC methods showed, on average, a 1.5X energy reduction and a bounded energy overhead, respectively. Furthermore, analysis of lifetime and bitflip tolerability indicated 4.9X and 216.3X lifetime enhancements over binary IMC and in-memory SC methods, respectively, along with substantial bitflip tolerance. Eventually, it can be concluded from the results that this work has great potential for performing complex arithmetic computations in memory and as such can effectively perform high-order polynomials and Maclaurin expansions of complex computations. It also shows potential benefits for applications requiring approximations to identify statistical anomalies in large datasets. In the future, we aim to leverage the proposed method to enable more applications to be effectively executed in memory by exploiting SC.